\documentclass[aps,prb,twocolumn,showpacs,floatfix,superscriptaddress]{revtex4-1}
\usepackage{graphicx,epsfig,dcolumn,amssymb,amsmath}
\usepackage{booktabs}
\usepackage{float}
\usepackage{color}

\begin{document}

\title{Quantum Transport Characteristics of Lateral pn-Junction of Single Layer TiS$_{3}$}

\author{F. Iyikanat}
\email{fadiliyikanat@iyte.edu.tr}
\affiliation{Department of Physics, Izmir Institute of Technology, 35430, Izmir, Turkey}

\author{R. T. Senger}
\affiliation{Department of Physics, Izmir Institute of Technology, 35430, Izmir, Turkey}

\author{F. M. Peeters}
\affiliation{Department of Physics, University of Antwerp, Groenenborgerlaan 171, 2020 Antwerp, Belgium}

\author{H. Sahin}
\email{hasansahin@iyte.edu.tr}
\affiliation{Department of Photonics, Izmir Institute of Technology, 35430, Izmir, Turkey}

\date{\today}
\date{\today}

\begin{abstract}

Using density functional theory and nonequilibrium Green's functions-based methods we 
investigated the electronic and transport properties of monolayer TiS$_{3}$ pn-junction. 
We constructed a lateral pn-junction in monolayer TiS$_{3}$ by using Li and F adatoms. An applied 
bias voltage caused significant variability in the electronic and transport properties of the TiS$_{3}$ 
pn-junction. In addition, spin dependent current-voltage characteristics of the constructed TiS$_{3}$ 
pn-junction were analyzed. Important device characteristics were found such as negative differential resistance 
and rectifying diode behaviors for spin-polarized currents in the TiS$_{3}$ pn-junction. These prominent 
conduction properties of TiS$_{3}$ pn-junction offer remarkable opportunities for the design of 
nanoelectronic devices based on a recently synthesized single-layered material.
\end{abstract}

\maketitle

\section{Introduction}
In recent years, two-dimensional materials have attracted a lot of 
interest due to their wealth of potential applications in various fields. Among the 
large family of two-dimensional materials, transition metal dichalcogenides (TMDs) 
stick out due to their exceptional electronic and optical properties.\cite{Novoselov, Wang, Chhowalla} 
TMDs are van der Waals (vdW) stacked layered materials.\cite{Zhao, Geim, Cunningham} 
Many of the TMDs have been shown to undergo an indirect-to-direct band gap transition when exfoliated 
down to a monolayer.\cite{Mak, Zhao1} Thus they are direct band gap semiconductor in the 
monolayer form. Due to the chemical versatility of this class of materials they exhibit a wide 
range of mechanical, electronic and optical characteristics.\cite{Jariwala, Splendiani, Manzeli, Elder}

In addition to TMDs there is important class of materials that are stable in atomic layer form:
transition metal trichalcogenides (TMTs). Similar to TMDs, TMTs also exhibit layered structures that are 
held together by weak vdW interactions. Most of the crystal structures of TMTs belong to the space 
group \textit{P2$_{1}$/m} and they consist of one-dimensional chains of trigonal prisms with the metal 
atom occupying the center of each prism. Unlike the TMDs, single, few layer or even macroscopicly thick TMTs 
may display direct-gap semiconducting behavior. Electronic and optical properties of various types of 
bulk TMTs such as ZrSe$_{3}$, HfSe$_{3}$, TiS$_{3}$, ZrS$_{3}$, ZrTe$_{3}$ have been investigated 
experimentally.\cite{Hoesch, Pacile, Gorlova1, Guilmeau} Previous studies have shown that, bulk TiS$_{3}$ is 
an n-type semiconductor with an energy band gap of 1 eV and it was shown that, it 
has a room temperature electronic mobility of about 30 cm$^2$ V$^{-1}$s$^{-1}$.\cite{Finkman} 
A nonlinear current-voltage characteristics of bulk TiS$_{3}$ has been observed below 60 \textit{K}.\cite{Gorlova2} 
Compared to TMDs, TMTs have drawn little attention until recently, which changed when the single layer 
TiS$_{3}$ was isolated.\cite{Island}

The exfoliation of a single layer of titanium trisulfide (TiS$_{3}$), has triggered 
tremendous interest in the electronic device community.\cite{Island}  Recently Jin \textit{et al.} 
theoretically studied structural, electronic and optical properties of titanium and 
zirconium trichalcogenide monolayers and they found that monolayers of TiTe$_{3}$ and 
ZrTe$_{3}$ are metallic, TiSe$_{3}$, ZrSe$_{3}$, and ZrS$_{3}$ are indirect while TiS$_{3}$ 
is a direct band gap semiconductor.\cite{Jin} Moreover, they showed that these monolayer 
semiconductors exhibit anisotropic conductive properties. The effect of tensile strain on 
the electronic properties of monolayers of TMTs have been studied.\cite{Li} We have recently 
investigated structural and electronic properties of defective and oxidized monolayer 
TiS$_{3}$.\cite{Iyikanat} Previous experimental studies on monolayer TiS$_{3}$ have revealed that it has a direct band 
gap of 1.10 eV and it exhibits a high carrier mobility in the few-layer limit.\cite{Ferrer, Ferrer1}. 
Recently, novel field-effect transistors based on few-layer TiS$_{3}$ were demonstrated.\cite{Lipatov, Island}. 
Island \textit{et al.} showed that, these transistors demonstrate promising device characteristics such as electron 
mobility of 2.6 cm$^{2}$ V$^{-1}$s$^{-1}$ and n-type semiconducting behavior with ultrahigh photoresponse 
and fast switching times at room temperature.\cite{Island} Strain engineering has been identified as one 
of the most promising strategies to tune the band gap because the reduced-dimensional structures 
can sustain much larger strains than bulk crystals.\cite{Kang1, Kang} Strong nonlinearity of the current-voltage 
characteristics has been reported by Gorlova \textit{et al.} \cite{Gorlova, Gorlova3} Especially, the nonlinearity in the 
current-voltage characteristics, the relatively high carrier mobility and the direct band gap properties even for a thickness 
of hundreds of layers make TiS$_{3}$ a potential candidate for future electronics.

In the present study, structural, electronic, magnetic and transport properties of p- and n-doped TiS$_{3}$ are 
investigated. The computational methodology is given in Sec. 2. Structural, electronic and magnetic properties of 
pristine, Li, and F doped monolayers of TiS$_{3}$ are studied in Sec. 3 and 4. Transport properties of TiS$_{3}$ 
pn-junction are investigated in Sec. 5. We summarize our results in Sec. 6.

\section{Computational Methodology}
All ionic and electronic relaxations are carried out within the framework of spin-polarized density functional theory 
(DFT) method implemented in the SIESTA package.\cite{Siesta} The Perdew-Burke-Ernzerhof generalized gradient approximation 
(PBE-GGA) is used for the exchange correlation functional. The Grimme's DFT-D2 dispersion correction is used to include 
the long-range vdW interactions. \cite{PBE, Grimme} A double-$\zeta$ polarized basis set is used and the mesh 
cutoff is chosen as 350 Ry. The convergence criterion for the energy is taken as 10$^{-5}$ eV. All atomic positions are 
fully relaxed with a force tolerance of 0.04 eV/\AA{}. A vacuum space of at least 15 \AA{} is used to prevent undesirable 
interactions between adjacent layers. Pressures on the unitcell are decreased to values less than 1.0 kBar. Structural 
calculations are performed with a $7\times11\times1$ Monkhorst-Pack Grid. A k-point sampling of $3\times3\times1$ is used 
for geometric calculations in $3\times4\times1$ supercell. The Bader charge population analysis is performed to calculate 
the charge transfer between adatoms and TiS$_{3}$.\cite{Henkelman}

The electronic transport calculations are performed by using the TRANSIESTA code, which is based on nonequilibrium Green's 
function formalism and DFT (NEGF-DFT).\cite{Transiesta} The current through the central region is evaluated by the 
Landauer-B\"{u}ttiker formula
\begin{equation}
I(V)=\frac{2e}{h}\int[f(E-\mu_{L})-f(E-\mu_{R})]T(E,V)dE,
\end{equation} 
where \textit{E} denotes the energy, $\mu_{L}$ and $\mu_{R}$ are the electrochemical potentials of the left and right electrodes, 
\textit{V} is the applied voltage, \textit{T(E, V)} is the quantum mechanical transmission probablity of electrons. \textit{e}, 
\textit{h} and \textit{f} denote electron charge, Planck's constant and Fermi function, respectively. Fully relaxed coordinates 
for the transport calculation, with a mesh cutoff of 350 Ry and a double-$\zeta$ polarized basis set are used. The Brillouin zone 
normal to the transport direction is sampled by a k-point mesh of $16\times1\times1$.

\begin{figure}
\includegraphics[width=8.5 cm]{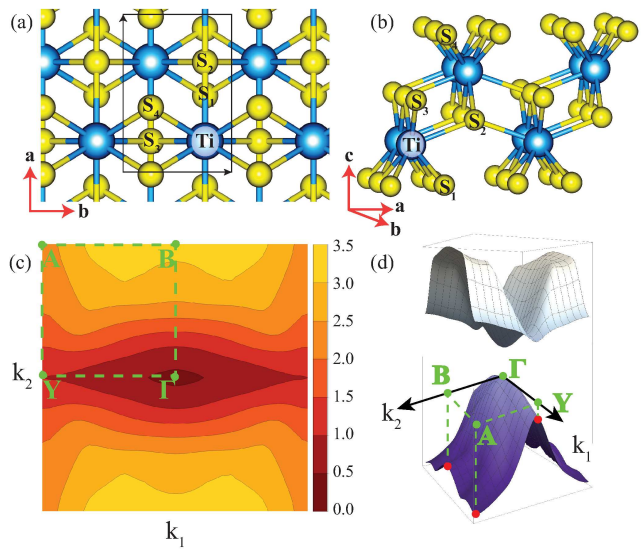}
\caption{\label{structure}
(Color online) 
(a) Top view of monolayer TiS$_{3}$ and its unit cell shown by a rectangle. For 
clarity, S atoms in the crystal are shown by yellow color and lowest layer S atom is 
presented as S$_{1}$, S atoms in the middle of the layer are shown by S$_{2}$ and S$_{3}$, 
S atom in the top layer is illustrated by S$_{4}$. Ti atom in the crystal is shown by blue color 
and labelled by Ti. (b) Side view geometry of TiS$_{3}$ along the \textit{a} direction. 
(c) Contour plots of the band gap (in eV). (d) Tilted side view of 
the three-dimensional surface plots of valence and conduction bands.}
\end{figure}

\section{Structural and Electronic Properties of Single Layer T\MakeLowercase{i}S$_{3}$}
The calculated lattice constants of monolayer TiS$_{3}$ are \textit{a} $ = 4.97$ \AA{} and \textit{b} $ = 3.40$ \AA{}, 
which are compatible with the previous studies. \cite{Kang, Jin} As seen in Fig. \ref{structure}(a), atoms in TiS$_{3}$ 
are arranged in a chain-like structure which consists of a prismatic trigonal structure with the metal atom located at 
the centre and chalchogen atoms located at the corners of the prism. The unit cell of TiS$_{3}$ consists of two Ti 
atoms and six S atoms, which is shown in Fig. \ref{structure}(a). To demonstrate more clearly the geometric structure of 
TiS$_{3}$, S atoms from different planes of TiS$_{3}$ are illustrated with different labels. Fig. \ref{structure}(b) 
presents a side view of TiS$_{3}$. As shown in this figure, S atoms located on the bottom plane of TiS$_{3}$ 
are labeled as S$_{1}$, whereas S atoms placed on the top plane of TiS$_{3}$ are labeled as S$_{4}$. These surface 
S atoms are bonded strongly to the Ti atoms with a bond lenght of 2.50 \AA{}. Correspondingly, S atoms located 
in the middle of the TiS$_{3}$ layer are denoted as S$_{2}$ and S$_{3}$. The locations of S$_{2}$ and S$_{3}$ atoms in the 
structure are geometrically equivalent, and the bond lenghts of the S$_{2}$ and S$_{3}$ atoms to the first and 
second nearest neighbour Ti atoms are 2.48 \AA{} and 2.65 \AA{}, respectively.

To investigate the electronic structure we performed a Bader analysis of the charge density as 
obtained from the SIESTA code. In a Bader charge analysis, atoms are defined by volumes bounded by surfaces of 
zero-flux. In these surfaces the gradient of the charge density is perpendicular to the normal vector of the surface. 
The integrated charges in the volume attributed to atoms are defined as Bader charge. In TiS$_{3}$, the bonds between 
inner-plane S atom (S$_{2}$ or S$_{3}$) and a Ti atom show ionic character with a charge transfer of 1.0 \textit{e} 
depleted from Ti atom to S atom. On the other hand, Ti atom donates 0.9 \textit{e} of valence charges to the two 
surface atoms (two S$_{1}$ or two S$_{4}$ atoms). These two surface S atoms share 0.9 \textit{e} and make a covalent 
bond with the Ti atom. As a result, in monolayer TiS$_{3}$, Ti atom has 2.1 \textit{e}, S$_{2}$ and S$_{3}$ atoms 
possess 7.0 \textit{e} and S$_{1}$ and S$_{4}$ atoms have 6.45 \textit{e}.

\begin{figure}
\includegraphics[width=8.5 cm]{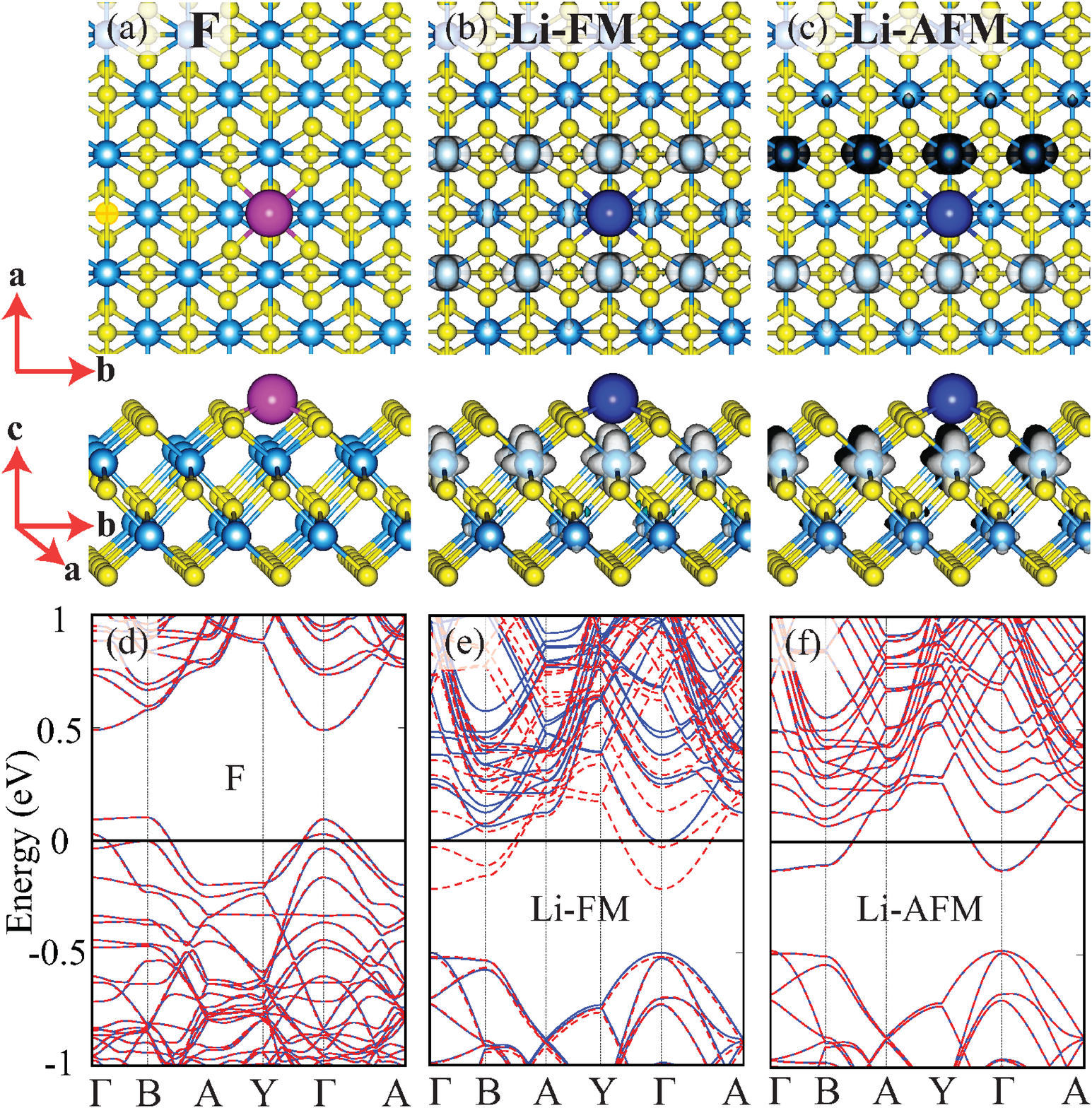}
\caption{\label{structure1}
(Color online) Top and side views of relaxed geometric structures and spin-dependent charge densities of 
TiS$_{3}$ are shown for F-doped layer (a), Li-doped layer in the ferromagnetic (Li-FM) (b), and antiferromagnetic (Li-AFM) 
(c) states. While white regions show majority spin charge densities, black regions show minority spin charge densities. 
Corresponding band structures of F-doped (d), Li-FM (e) and Li-AFM (f). Red dashed and blue solid lines show majority 
and minority spin bands, respectively. }
\end{figure}

Nanosheets and nanoribbons of TiS$_{3}$ are used to fabricate FETs. While fabricated nanosheets exhibit high mobilities, 
nanoribbons exhibit higher optical response and gate tunability, but have lower mobility. \cite{Island} Previous experimental 
and theoretical studies have shown that, exceptional chain-like structure of TiS$_{3}$ leads to anisotropic conductive 
properties, i. e. the resistivity of TiS$_{3}$ across the chains are significantly higher than the resistivity along 
the chains. \cite{Gorlova1, Jin} In order to properly understand the electronic properties of TiS$_{3}$, contours 
plots of band gap energies in reciprocal space of TiS$_{3}$ are depicted in Fig. \ref{structure}(c). Monolayer TiS$_{3}$ 
is a direct-gap semiconductor with a band gap of 1.10 eV.\cite{Iyikanat, Li, Jin}. Our PBE calculated band gap value of 
TiS$_{3}$ is 0.45 eV. It is well known that the bare PBE underestimates the band gap values of semiconductors. However, 
the electronic band dispersion is well-approximated by PBE. As seen from Fig. \ref{structure}(c), TiS$_{3}$ has a direct 
band gap with both the valence-band-maximum (VBM) and the conduction-band-minimum (CBM) residing at the \textit{$\Gamma$} 
point, and variation of the gap along the \textit{$\Gamma$}-\textit{Y} direction is small. VB and CB give quite valuable 
information about the electronic structure of a material. Three-dimensional surface plots of VB and CB of TiS$_{3}$ are 
shown in Fig. \ref{structure}(d). The plotted surfaces clearly show that VB and CB surfaces are highly anisotropic. 
The electron mobility along k$_{2}$ direction is much larger than that along the perpendicular direction. This makes it 
clear that the contribution to the electronic conductivity parallel to the chain direction will be much larger than 
that corresponding to normal to the chain direction. 

\section{p- and n- Type Doping of T\MakeLowercase{i}S$_{3}$}

Top and side views of the relaxed geometric structure of F-doped TiS$_{3}$ are shown in Fig. \ref{structure1}(a). 
Li-doped TiS$_{3}$ has magnetically two different states which are denoted as Li-FM and Li-AFM and shown in 
Figs. \ref{structure1}(b) and (c), respectively. The structural and electronic properties of these states are given 
in Table \ref{el_pro}. As shown in Figs. \ref{structure1}(a)-(c), both adatoms are placed in the middle of 
the four S atoms. While Li atom binds at a distance 0.86 \AA{}, F atom is located 0.79 \AA{} above the surface. In order 
to hinder the interactions between the adatoms of adjacent supercells, a $3\times4\times1$ supercell is used for electronic 
structure calculations. Both Li and F atoms do not cause any significant structural deformation, thus TiS$_{3}$ preserves 
its lattice constant values, which are in the supercell form $a = 14.90$ and $b = 13.61$ \AA{}. 
The binding energies are calculated using the expression $E_{bind} = E_{TiS_{3}+Adatom} - E_{TiS_{3}} - E_{Adatom}$, 
where \textit{E$_{bind}$} is the binding energy of the considered adatom, \textit{E$_{TiS_{3}+Adatom}$} is the energy of adatom 
doped TiS$_{3}$, \textit{E$_{TiS_{3}}$} is the energy of pristine TiS$_{3}$ and \textit{E$_{Adatom}$} is the energy of the considered 
adatom. As listed in Table \ref{el_pro}, binding energies of F, Li-FM, and Li-AFM on TiS$_{3}$ are calculated to be 2.48 and 
2.98, and 2.99 eV, respectively. 

\begin{table}[htbp]
\caption{\label{el_pro}Perpendicular distance of the adatom from the surface of the layer, $h$. 
The preferred binding site, where the hollow site is denoted by H. The binding energy of adatom, 
E$_{bind}$. Bader charge transferred from adatom to TiS$_{3}$, $\Delta\rho$. Magnetic moment 
per supercell, $\mu$.}
\begin{tabular}{lccccccc}
\hline\hline
      & $h$     &Binding&   E$_{bind}$ & $\Delta\rho$  & $\mu$\\
    & (\AA{}) &Site   &   (eV)       &($\textit{e}$)&($\mu_{B}$)\\
\hline
F       & 0.79 &  H   &    2.48       &-0.6& 0.0   \\
Li-FM      & 0.86 &  H   &    2.98        &+1.0& 0.9  \\ 
Li-AFM     & 0.86 &  H   &    2.99        &+1.0& 0.0  \\ 
\hline\hline 
\end{tabular}
\end{table}

Bader charge analysis shows that F adatom gains 0.6 \textit{e} when it is placed on TiS$_{3}$. Contrary to F, 
Li adatom donates all its valence charge (1 \textit{e}) to TiS$_{3}$. The non-magnetic nature of the monolayer 
TiS$_{3}$ is not affected by F doping. However, Li doping significantly changes the magnetic properties 
of TiS$_{3}$ and two different magnetic states are generated. One of them is denoted as Li-FM state, and the other 
is denoted as Li-AFM state and as given in Table \ref{el_pro} their magnetic moment values are 0.9 and 0.0 $\mu_{B}$ 
per supercells. Our results show that Li-FM is the ground state. However, calculated energy difference between 
Li-FM and Li-AFM states is only 5 meV. Thus, Li-AFM state can not be ignored and both of the states are considered in 
the present paper. Spin polarized charge densities of Li-doped TiS$_{3}$ for FM and AFM states are shown from top 
and side views in Figs. \ref{structure1}(b) and (c), respectively. As seen in the figure of Li-FM, Ti atoms of the nearest two 
chains to the Li adatom are polarized ferromagnetically. While the nearest Ti atom to the Li adatom has a magnetic 
moment of 0.21 $\mu_{B}$, this value decreases to 0.14 $\mu_{B}$ for the furthest Ti atom. 
On the other hand, Fig \ref{structure1}(c) shows that Ti atoms of the nearest two chains to the Li adatom 
are polarized antiferromagnetically. As in the case of Li-FM, magnetic moments of individual Ti atoms are decreasing 
when the distance to the Li adatom is increased. Absolute values of magnetic moment of individual Ti atoms are almost the same 
in the Li-FM and Li-AFM states. However, in both cases S atoms do not exhibit any significant magnetic 
moment.

In order to investigate the effect of adatom doping on the electronic structure of TiS$_{3}$, band diagrams of 
F- and Li-doped $3\times4\times1$ TiS$_{3}$ are calculated and showed in Figs. \ref{structure1}(d)-(f). As mentioned above 
Li-doped TiS$_{3}$ has two different magnetic states with very similar energy values and  their band diagrams are also 
shown in these figures. Fig. \ref{structure1}(d) shows that, when TiS$_{3}$ is doped with F atoms, some states of TiS$_{3}$ discharge 
because F atom recieve 0.6 \textit{e$^{-}$} from it. Thus, F doping results in acceptor states close to the top of the valence 
band. These results suggest that F doping causes p-type doping of TiS$_{3}$. Flat bands of F atoms appear $\sim$ 0.3 
eV below the Fermi level. On the other hand, donor states occur close to the conduction band minimum when TiS$_{3}$ is 
doped with Li atoms. Fig. \ref{structure1}(e) shows that, the donor states come from only the majority spin component in 
the Li-FM state. However, as shown in Fig. \ref{structure1}(f), in the Li-AFM state all bands are degenerate 
and the donor states originate from charges with both spin components. Adsorption of Li atoms on TiS$_{3}$ leads to n-type 
doping. Created donor and acceptor states are necessary for the construction of a pn-junction. Thus, doping one side of a 
TiS$_{3}$ layer with F (a p-type dopant) and the other side with Li (an n-type dopant) forms a pn-junction. 
One can modulate the electronic properties of TiS$_{3}$ by changing the atom type of the dopant and the doping 
concentrations.

\begin{figure}
\includegraphics[width=8.5 cm]{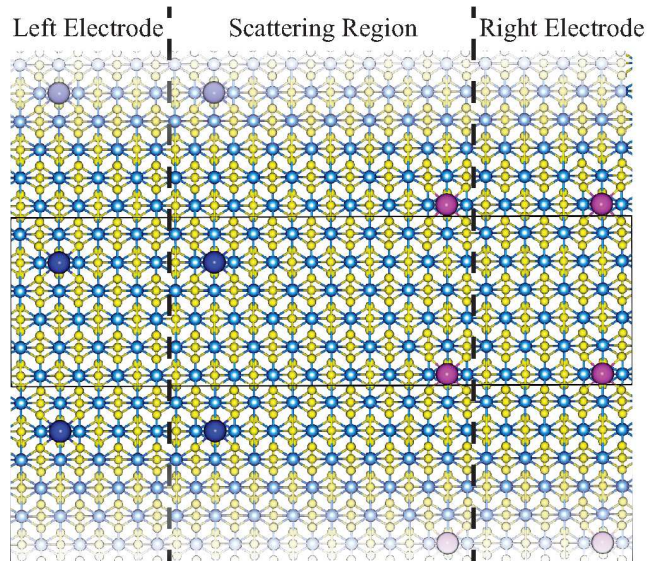}
\caption{\label{label_3}
(Color online) Li and F doped device structure of TiS$_{3}$ pn-junction. Blue and purple atoms 
illustrate Li and F atoms, respectively.
}
\end{figure}

\section{Transport properties of T\MakeLowercase{i}S$_{3}$ PN-JUnction}

A pn-junction can be constructed by combining p-type and n-type materials in lateral or vertical 
orientation. Because of the controllability of its electronic properties and its possible 
applications, pn-junctions are used in many semiconductor devices. To construct a pn-junction, the 
presence of shallow p- and n-type doping levels are essential. In the previous part, 
we have shown that, p- and n-type doping of TiS$_{3}$ can be realized through doping with F and Li atoms, 
respectively. Thus when we construct a device which consist of p-type and n-type doped parts of TiS$_{3}$, 
states near the valence and conduction bands will behave like acceptor and donor states. The quantum transport 
properties of a monolayer TiS$_{3}$-based pn-junction are investigated in the following.

The device setup used for the transport calculations on the TiS$_{3}$ pn-junction is shown in Fig. \ref{label_3}. 
The device geometry consists of semi-infinite left electrode, semi-infinite right electrode, and a scattering 
region. The possibility of experimental realization of TiS$_{3}$ pn-junction is higly dependent on 
the dopant atom concentrations. Thus, to simulate the electrode parts of the device, one adatom of Li or F 
and a $3\times1\times4$ supercell of TiS$_{3}$ are used. In the present study, ground state of Li-FM is used for the transport 
calculations of TiS$_{3}$ pn-junction. The adatom concentration is the same $\sim$ 5$\times$10$^{13}$ 
cm$^{-2}$ for both electrodes. The distance between Li and F atoms in the scattering region is taken to be 22.7 \AA{}. 
Recent studies have shown that, monolayer TiS$_{3}$ exhibits conductance anisotropy, and the electron mobility along the 
chain direction is larger than across the chain direction.\cite{Gorlova1, Jin} Therefore, the transport direction 
of the device geometry is choosen to be the chain direction.

\begin{figure}[H]
\includegraphics[width=8.5 cm]{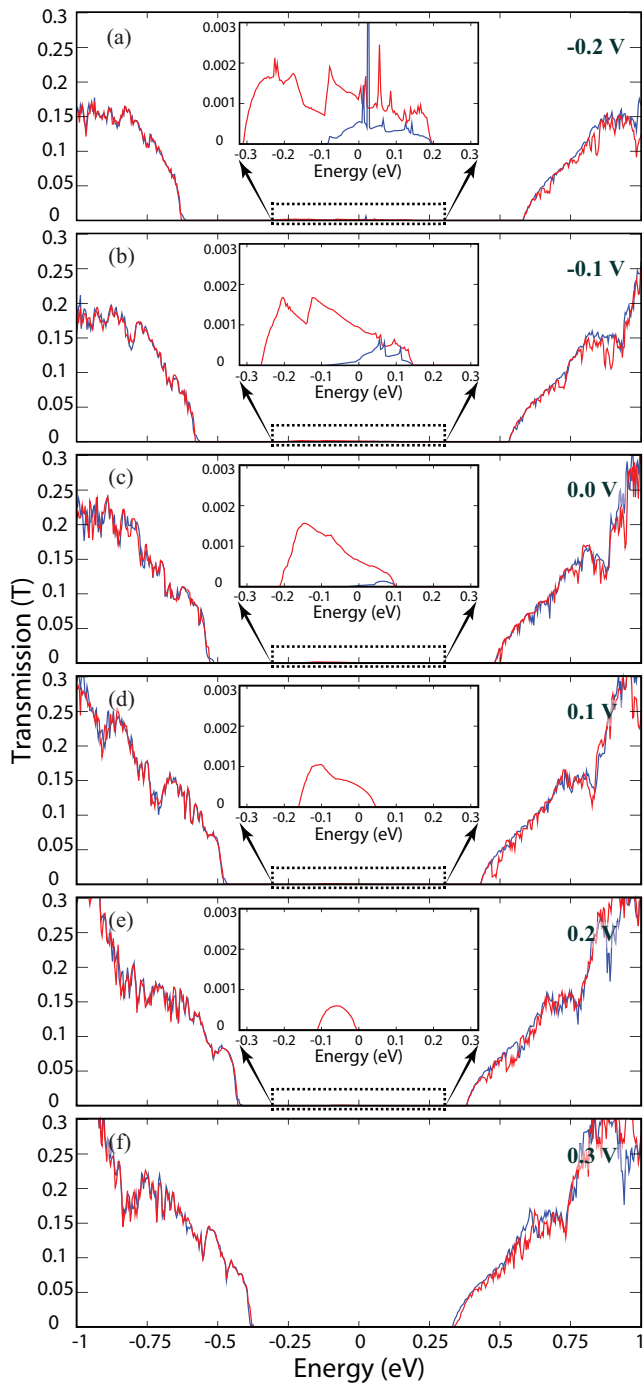}
\caption{\label{label_4}
(Color online) Transmission of TiS$_{3}$ pn-junction for different bias. Insets are zoomed 
transmissions in the vicinity of Fermi level. Transmission near the Fermi level completely 
diminish for bias of 0.3 V. Red and blue lines show majority and minority spin components 
of the transmission profiles, respectively.}
\end{figure}

Fig. \ref{label_4} presents transmission profiles per unit lenght of constructed TiS$_{3}$ pn-junction 
for varying bias voltages. As shown in Fig. \ref{label_4}(c), in the unbiased situation 
TiS$_{3}$ pn-junction has a transport gap of 0.99 eV. VBM and CBM of this transmission gap 
coincides with the VBM of Li doped TiS$_{3}$ and CBM of F doped TiS$_{3}$, respectively. Inset of 
Fig. \ref{label_4}(c) shows transmission profile of TiS$_{3}$ pn-junction zoomed around 
the Fermi level, which is shown by dashed rectangle. These small transmissions occur when the valence 
band states of F doped region overlap with the conduction band states of Li doped region. Therefore, 
charges from valence band of the F doped source region can tunnel to the empty conduction band of Li 
doped region. As shown in Figs. \ref{label_4} (a)-(f), applying bias voltage leads to significant modulation in 
the electronic properties of TiS$_{3}$ pn-junction. 

First of all, increasing bias voltage of pn-junction shifts the energy bands of F doped 
electrode to the lower energy values, whereas it shifts the energy bands of Li doped electrode 
to higher energy values. As a result, when the bias voltage is increased from -0.2 V to 0.3 V, 
the transport gap of the TiS$_{3}$ pn-junction decreases from 1.19 eV to 0.68 eV. 

Secondly, while transmission values near to the VBM and CBM look similar, for high energy values such as 
-1.0 eV, transmission values are increasing from 0.15 to 0.30 with increasing bias voltage from -0.2 V to 
0.3 V. As can be seen from Fig. \ref{structure1}, near to the edges of the CBM of F doped TiS$_{3}$ and 
the VBM of Li doped TiS$_{3}$ consist of a smaller number of bands relative to the deeper energies. When 
bias voltage is increased, deeper states of p-type doped and n-type doped parts of the TiS$_{3}$ contribute 
to the transmission. Thus, rising of the transmission values of TiS$_{3}$ pn-junction to high energy ranges 
can be expressed by the shifting of deeper states of the left and right electrodes. On the other hand, spin-up 
and spin-down transmission profiles are almost the same for the considered energy range and bias values. 

Lastly, Fig. \ref{label_4} shows that, transmission in the vicinity of the Fermi level is decreasing 
with increasing bias voltage from -0.2 to 0.3 V. Due to quantum tunnelling between the electrodes, there is 
only small transmission between energy values of -0.2 and 0.1 eV, even in the unbiased situation, which 
is shown in the inset of Fig. \ref{label_4}(c). In the previous part, we showed that Li doped 
TiS$_{3}$ is magnetic. Ti atoms near to the adsorbed atom possess a net magnetic moment 
with the same spin polarization. Therefore, transmission values of TiS$_{3}$ pn-junction near the Fermi 
level are highly spin-polarized. Figs. \ref{label_4}(d)-(f) show that, with increasing bias voltage, the transmission near 
the Fermi level decreases and at 0.3 V bias it is completely suppressed. On the other hand, when the bias voltage 
is increased in the other direction, energy range with the nonzero transmission and transmission 
values near the Fermi level increases for both spin components. Moreover, some small resonant 
states occur for applied bias voltage of -0.2 V, which is shown in the inset of Fig. \ref{label_4}(a).

\begin{figure}
\includegraphics[width=6.5 cm]{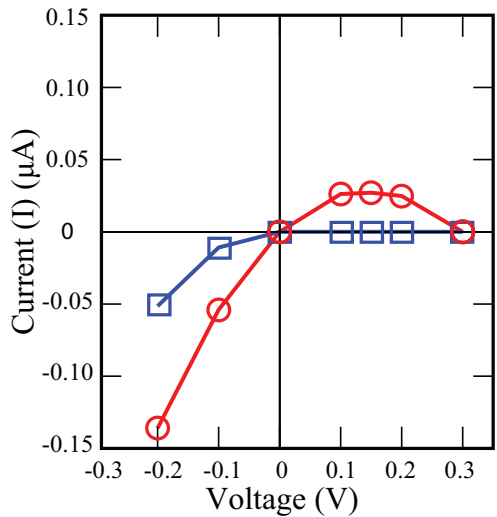}
\caption{\label{label_5}
(Color online) The calculated spin-up (red curve) and spin-down (blue curve) current-voltage (\textit{I-V}) 
characteristics for TiS$_{3}$ pn-junction.}
\end{figure}

The spin-up and spin-down \textit{I-V} characteristics of TiS$_{3}$ pn-junction are shown in Fig. \ref{label_5} 
by the red and blue curves, respectively. The corresponding current values are calculated for a
scattering region of size 14.91 \AA{}. Fig. \ref{label_5} explicitly shows that, spin-up and spin-down \textit{I-V} 
characteristics of TiS$_{3}$ pn-junction are rather different. Also this figure clearly shows that 
spin-up current exhibits negative differential resistance (NDR), when positive bias applied to the system. 
Spin-up and spin-down current show typical diode behavior for negative bias. The NDR phenomenon in this 
system can be explained by the variation of transmissions and the relative shifts of energy states of device and 
electrode regions with applied bias voltage. As shown in Figs. \ref{label_4}(c)-(e), small spin-up transmission 
peaks near the Fermi level are visible for 0.0 $\leq$ \textit{V} $<$ 0.3, and these transmission peaks disappear for bias 
voltage 0.3 V, whereas spin-down transmissions are zero for these positive bias voltage values. These small spin-up 
transmission peaks lead to negative differential resistance (NDR) between 0.0 V and 0.3 V in the spin-up \textit{I-V} 
characteristic of TiS$_{3}$ pn-junction, as shown in Fig. \ref{label_5}. However, the spin-down current is zero for 
the positive bias voltage. On the other hand, when negative bias voltage is applied, spin-up and spin-down transmission 
peaks occur near the Fermi level, with the dominant one being spin-up, as shown in Figs. \ref{label_4}(a), (b). As a result, 
the spin-up electrons flow more easily from the valence band of the F-doped electrode to the conduction band of the Li-doped 
electrode, than the spin-down electrons. This explains the higher spin-up current values of the TiS$_{3}$ pn-junction for 
negative bias. Moreover, spin-up and spin-down transmission peaks near the Fermi level rise with increasing negative bias 
voltage. This leads to an increase in the spin-up and spin-down current values for higher negative bias voltages. 
As a result, the \textit{I-V} characteristic of a TiS$_{3}$ pn-junction is strongly related to the doping types, 
its effect on the magnetism and the amount of doping of the electrodes and scattering parts. While 
one spin component of the current shows NDR, both spin components of the current show a typical diode character. 

\section{Conclusions}
In summary, we investigated electronic and transport properties of monolayer TiS$_{3}$ pn-junction by 
using density functional theory and nonequilibrium Green's functions method. Doping one side of a TiS$_{3}$ 
layer with F adatom (a p-type dopant) and the other side with Li adatom (an n-type dopant) forms a lateral 
pn-junction. We showed that the conduction properties of TiS$_{3}$ pn-junction can be tuned by applying a 
bias voltage. The predicted device showed NDR behavior only for one spin component of the current. Both spin 
components of the current had asymmetric characteristics with respect to the bias voltage, thus, they showed 
rectifying diode behavior. Prediction of a pn-junction based on monolayer TiS$_{3}$ reveals a functional 
application of this new material.

\section{acknowledgments}
This work was supported by the bilateral project between TUBITAK (through Grant No. 113T050) and
the Flemish Science Foundation (FWO-Vl). The calculations were performed at TUBITAK ULAKBIM, High
Performance and Grid Computing Center (TR-Grid e-Infrastructure). FI, HS, and RTS acknowledge the 
support from TUBITAK Project No 114F397. H.S. acknowledges support from Bilim Akademisi-The Science
Academy, Turkey under the BAGEP program.

\end{document}